\documentclass{article} 
\usepackage{spconf,amsmath,epsfig}

\usepackage{graphicx}
\usepackage{xcolor}
\usepackage{wrapfig}
\usepackage{multirow}

\usepackage{float}
\floatstyle{plaintop}
\restylefloat{table}

\title{Sentinel-1 SAR based weakly supervised learning for tropical forest mapping}

\name{Adugna Mullissa${^{1,}}{^2}$,  Sassan Saatchi ${^{1,}}{^{2,}}{^3}$\thanks{}}
\address{${^1}$ CTREES.org, Pasadena, CA 91105, United States \\
${^2}$ UCLA, Institute of Environment and Sustainability, Los Angeles, CA 90095, United States \\
${^3}$ NASA, Jet Propulsion Laboratory, Pasadena, CA 91109, United States \\
amullissa@ucla.edu}

\begin{document}

\maketitle
\begin{abstract}
\noindent Tropical forests play an important role in regulating the global carbon cycle and are crucial for maintaining the tropical forest biodiversity. Therefore, there is an urgent need to map the extent of tropical forest ecosystems. Recently, deep learning has come out as a powerful tool to map these ecosystems with the caveat of curating high quality reference datasets. Since, manually annotating high quality reference datasets is time consuming and expensive, weakly supervised learning techniques offer the potential to train high quality models without the need for manually annotating large quantities of reference datasets. In this manuscript, we propose two weakly supervised approaches that are based on Sentinel-1 SAR images, sparsely distributed pixel-wise high quality reference labels and densely distributed noisy reference labels. The proposed approaches were tested in a tropical setting in the Brazilian amazon. The results demonstrate that high quality tropical forest maps can be derived from weakly supervised learning without the need for manually annotated labels.  
\end{abstract}

\section{Introduction}
\label{sec:intro}

Tropical forests play an important role in regulating the global carbon and water cycle and provides bio-diversity and ecosystem services \cite{lewis2009changing}. Therefore, mapping tropical forest helps in the quantification of carbon stocks to mitigate climate change and also enables timely interventions such as policy or law enforcement response to illegal deforestation. For this purpose, optical satellite images were widely used to map tropical forests but were often limited by pervasive cloud covers existing in these regions. In this regard, the advent of Synthetic Aperture Radar (SAR) sensors was a game changer as it allowed the acquisition of images in almost all weather, enabling the mapping of forests in any locations irrespective of cloud cover.\\

\begin{figure*}
	\centering
	\begin{tabular}{cc}
		\includegraphics[width=0.5\textwidth]{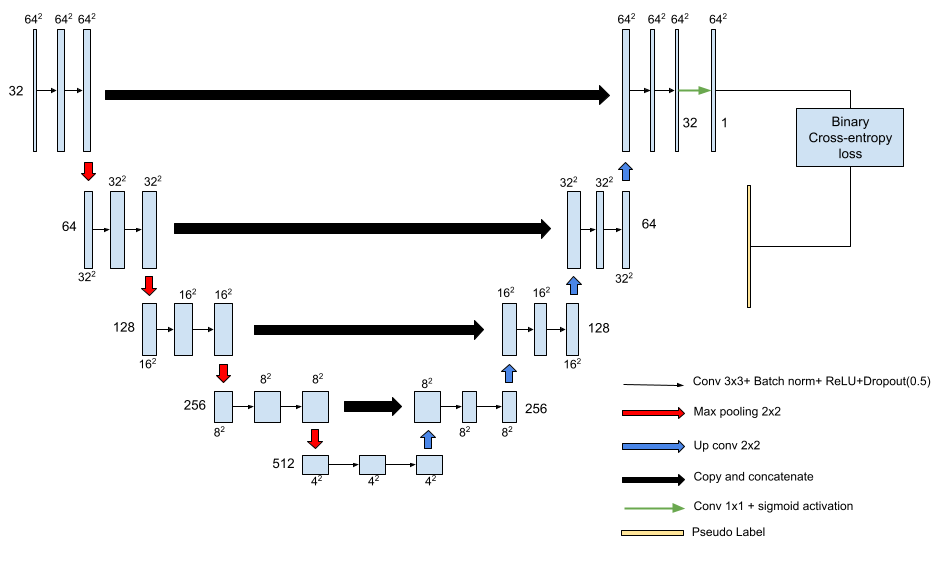}&
        \includegraphics[width=0.5\textwidth]{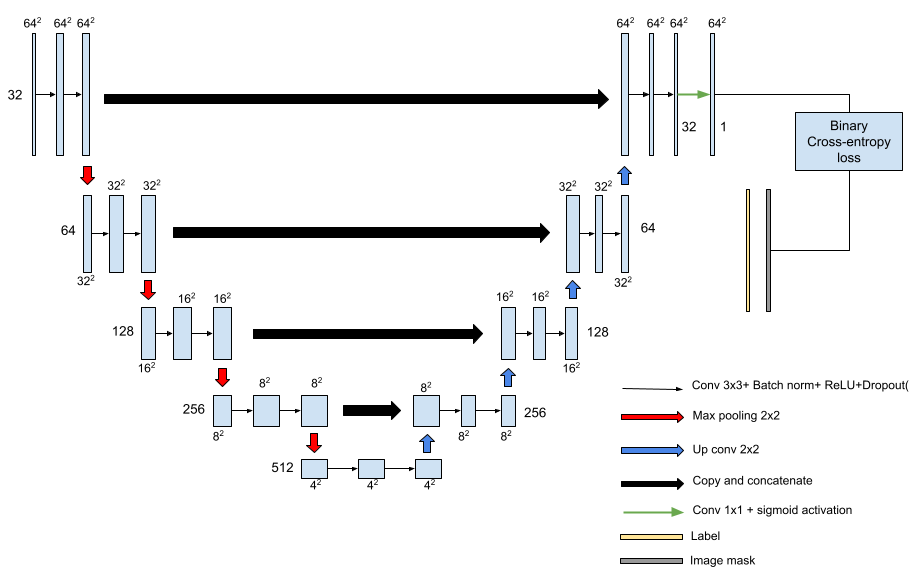}\\
		{(a)}&{(b)} \\
	\end{tabular}
	\caption{The Unet-like architecture used in the proposed method.  (a) In the inaccurate supervision approach, we synthesize a low quality pseudo-label from globally available coarse resolution Hansen global forest cover/ forest loss dataset \cite{hansen2013high} in the first step to train a Unet-like model and will use the prediction from the model to use as a label in subsequent steps. (b) The incomplete supervision approach, we use high quality pixels that are synthesized by masking 98\% of the labels in the image patches to train a Unet-like model.    }\label{fig:Fig1}
\end{figure*}

\noindent The recent success of deep learning in computer vision proliferated its application in the field of remote sensing for the purpose of landcover mapping \cite{kussul2017deep}. In particular, deep fully convolutional neural networks such as Unet \cite{ronneberger2015u} has been ubiquitous in remote sensing based semantic segmentation tasks. However, one of the main challenges in applying these deep learning architectures was the lack of high quality reference data that are required for training. In their absence, users often resort to manual annotation to synthesize high quality reference data. Recently,  weakly supervised learning (WSL) approaches were adopted to train a high quality model without the need to annotate large quantities of reference datasets. Weakly supervised machine learning approaches are techniques that use higher level and often noisier labels to synthesize a large quantity of reference labels by iterative refinement \cite{zhou2018brief}. WSL are often categorized into three groups as incomplete supervision, inexact supervision or inaccurate supervision \cite{mullissa2023deep}. Incomplete supervision refers to the case when the available reference labels are precise but few in number to train a high quality model and there exists sufficient supply of unlabeled data. On the other hand, inexact supervision happens when the labelling is coarse or high level such as image level annotations whereas the required level is pixel level annotations. Inaccurate supervision indicates the reference labels are crowd-sourced datasets that are often noisy, thereby requiring iterative refinement before application. \\

\noindent A few notable works applying WSL to remote sensing data include the work by the authors in \cite{wang2020weakly}  to apply class activation maps and masked Unet approach to segment croplands. The authors in \cite{li2023one} used image level labels in the inexact supervision category for the multi-class landcover segmentation of aerial and satellite images. Whereas, the authors in \cite{li2021effectiveness} demonstrated the effectiveness of WSL using high resolution satellite images for building extraction. \\

\noindent In this manuscript, we demonstrate and compare two weakly supervised approaches based on the incomplete supervision and inaccurate supervision to derive a forest non-forest cover map in the Brazilian Amazon using Sentinel-1 SAR images.\\  

\section{Methodology}
\label{sec:intro}

\subsection{Incomplete supervision}

\noindent We use a sparse forest/non-forest pixels that were extracted from the intersection between the JAXA ALOS-PALSAR2 based global four class forest/non-forest map \cite{shimada2014new}, Hansen global forest cover/ forest change dataset \cite{hansen2013high} and the Land Use Change Alerts \cite{mullissa2024luca} recently released to the public through an interactive app. We apply a masked approach to randomly extract 2\% of pixels in the reference label patch to synthesize the reference labels. The extracted labels were visually verified with a Planet scope image before training a Unet-like model for semantic segmentation of Sentinel-1 SAR images  (Figure~\ref{fig:Fig1}b) to map forest cover.  In this way, we force the network to learn from the 2\% of pixels available within the patch.  Therefore, the deep learning objective can be formulated as:

\begin{equation}
\displaystyle
 \mathrm{arg}\min_{\theta} \sum  L(f_{\theta}(S_1), {R}).
\end{equation}

\begin{figure*}
	\centering
	\begin{tabular}{cccc}
		\includegraphics[width=0.3\textwidth]{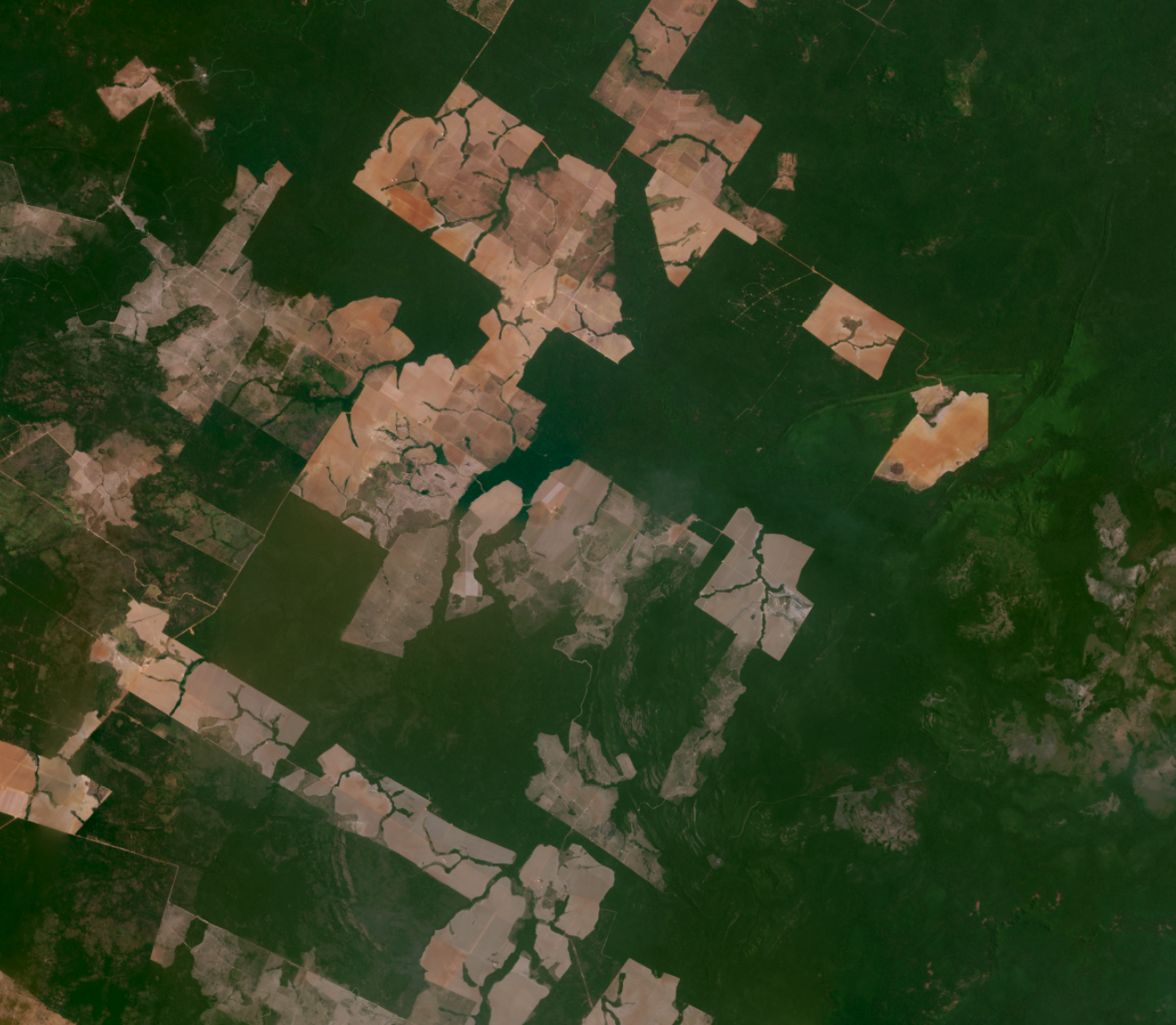}&
		\includegraphics[width=0.3\textwidth]{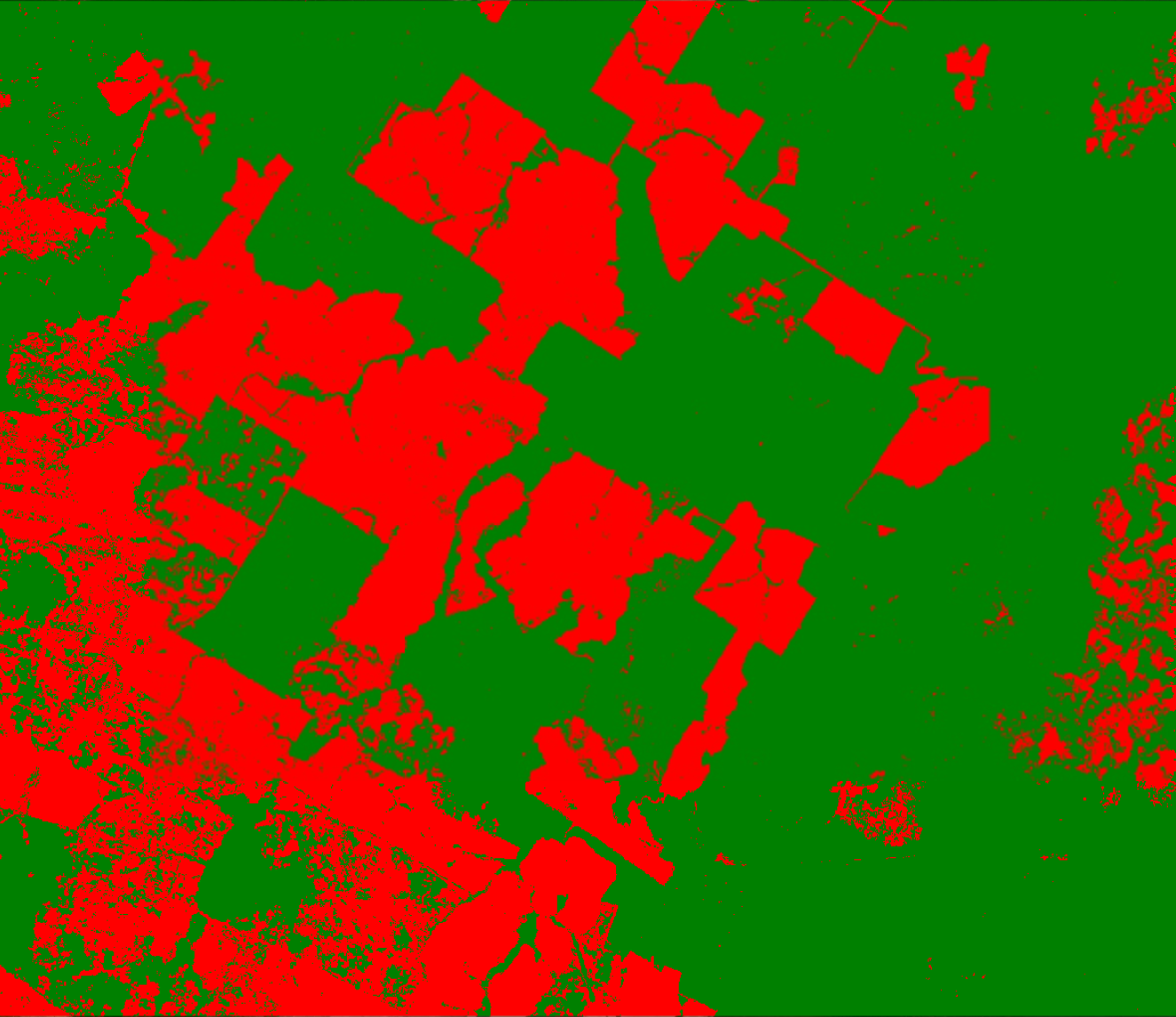}&
        \includegraphics[width=0.3\textwidth]{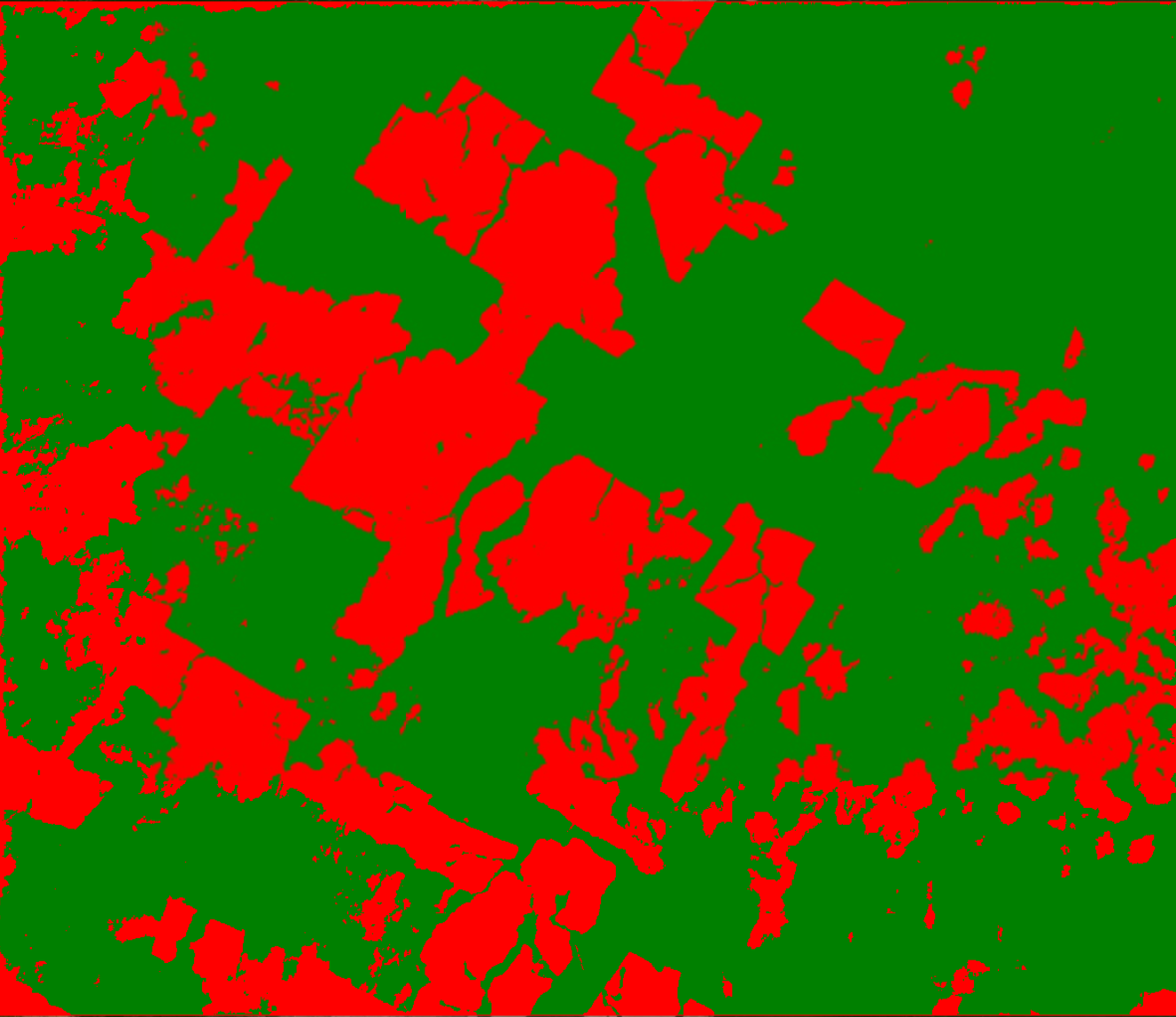}\\
		{(a)}&{(b)}&{(c)} \\
	\end{tabular}
 	\begin{tabular}{c}
\includegraphics[width=0.2\textwidth]{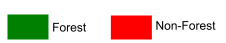}\\

	\end{tabular}
	\caption{The reference labels that are iteratively refined using the inaccurate supervision weakly supervised approach (a) Planet scope image acquired over the training area. (b)  The low quality reference data synthesized from the Hansen global forest cover dataset that was used as an initial pseudo-label. (c) the final iteratively refined reference label.   }\label{fig:Fig2}
\end{figure*}

\noindent Here, $S_1$ are the input Sentinel-1 SAR image,  $R$ is the available 2\% binary forest non-forest pixel labels we used as reference data  and $f_\theta$ is the deep learning network parameterized by the learned parameters $\theta$.  The Loss function (L) in (1) is a binary cross-entropy loss function. \\

\subsection{Inaccurate supervision}

\noindent For the inaccurate supervision approach, we use a globally available coarse resolution and noisy forest/non-forest map provided by \cite{hansen2013high}. We use these complete albeit inaccurate reference maps as pseudo-labels to train an initial model. We used the same Unet-like architecture as the incomplete supervision to train the network. We use the prediction from the initial model to use as a pseudo-label to train a subsequent Unet-like model (Figure~\ref{fig:Fig1}a).  We continue this sequence until number of pixels in the newly predicted forest/non-forest maps class changes is below 10\% compared to a previous iteration.  \\ 

\subsection{Architecture}

\noindent The Unet-like architecture consisted of an encoder block that downsamples the feature maps to its latent space and a decoder block that reconstructed the feature maps to their original sizes. The architecture consisted of 5 encoder blocks each consisting of of 2 convolution layers interleaved with batch normalization, rectified linear unit activation (ReLU), pooling layer and 50\% dropout layer. Whereas the decoder block consists of transpose convolutions interleaved with batch normalization and ReLU activation layers. Residual connections connect the encoder and decoder layers at each level. Finally a segmentation head with $1\times1$ convolution and a Sigmoid activation provides the probability of forest (Figure~\ref{fig:Fig1}). The architecture used in this manuscript is implemented using the TensorFlow deep learning library in a Python 3.10 environment and trained on the Amazon Web Services (AWS) platform using 4 Tesla T4 GPU's.\\

 \begin{figure*}
	\centering
	\begin{tabular}{cccc}
		\includegraphics[width=0.22\textwidth]{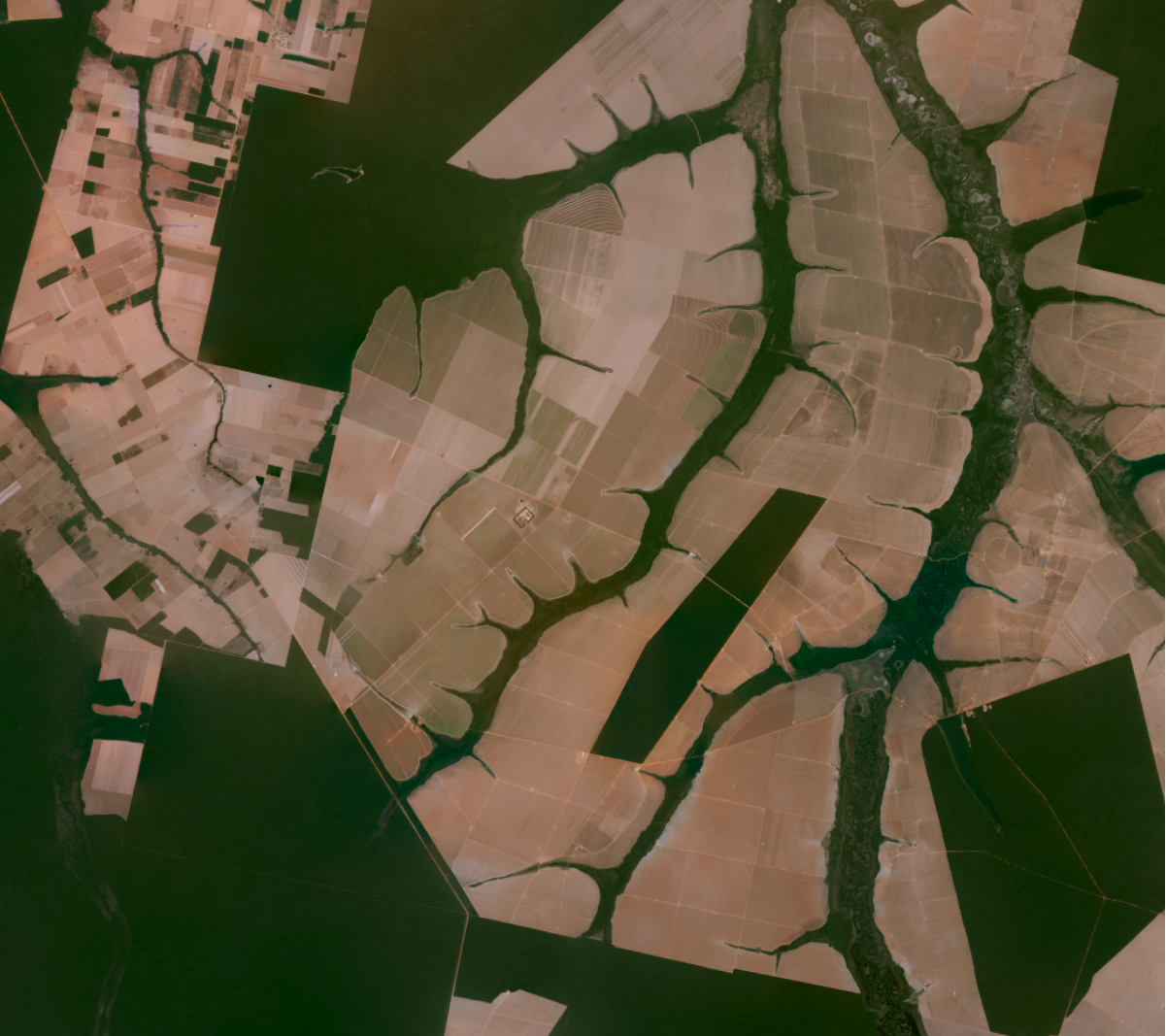}&
        \includegraphics[width=0.22\textwidth]{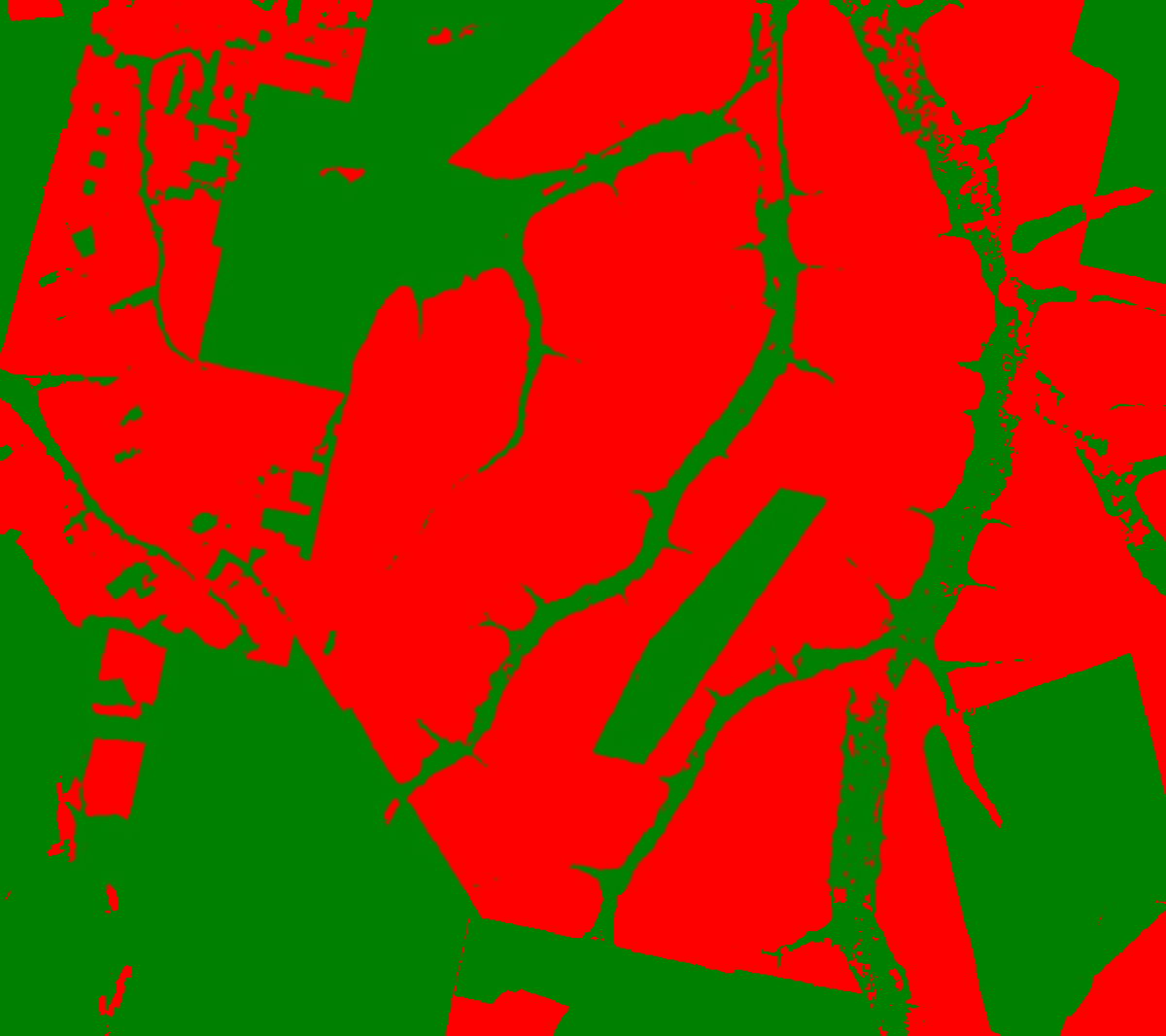}&
		\includegraphics[width=0.22\textwidth]{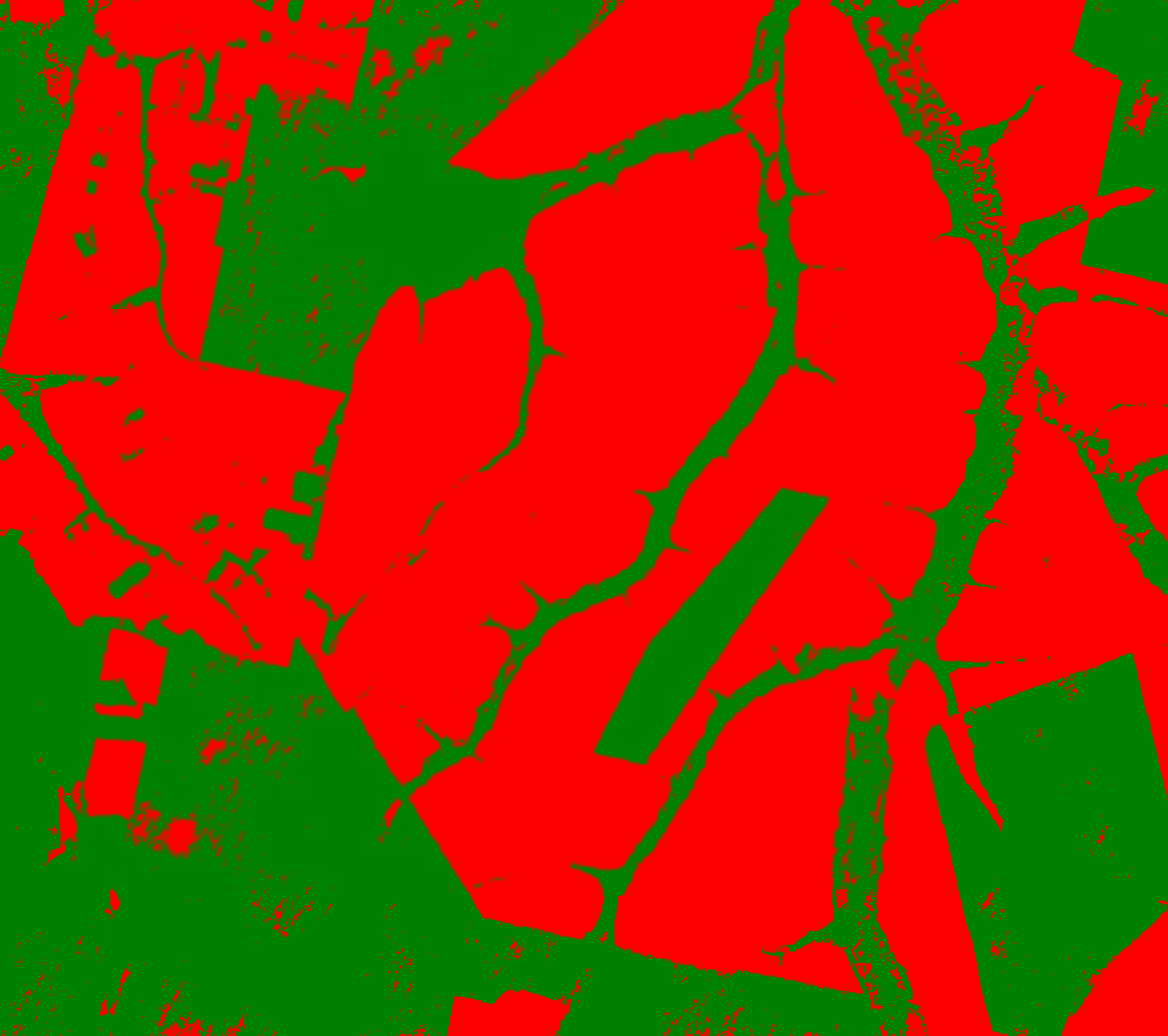}&
        \includegraphics[width=0.22\textwidth]{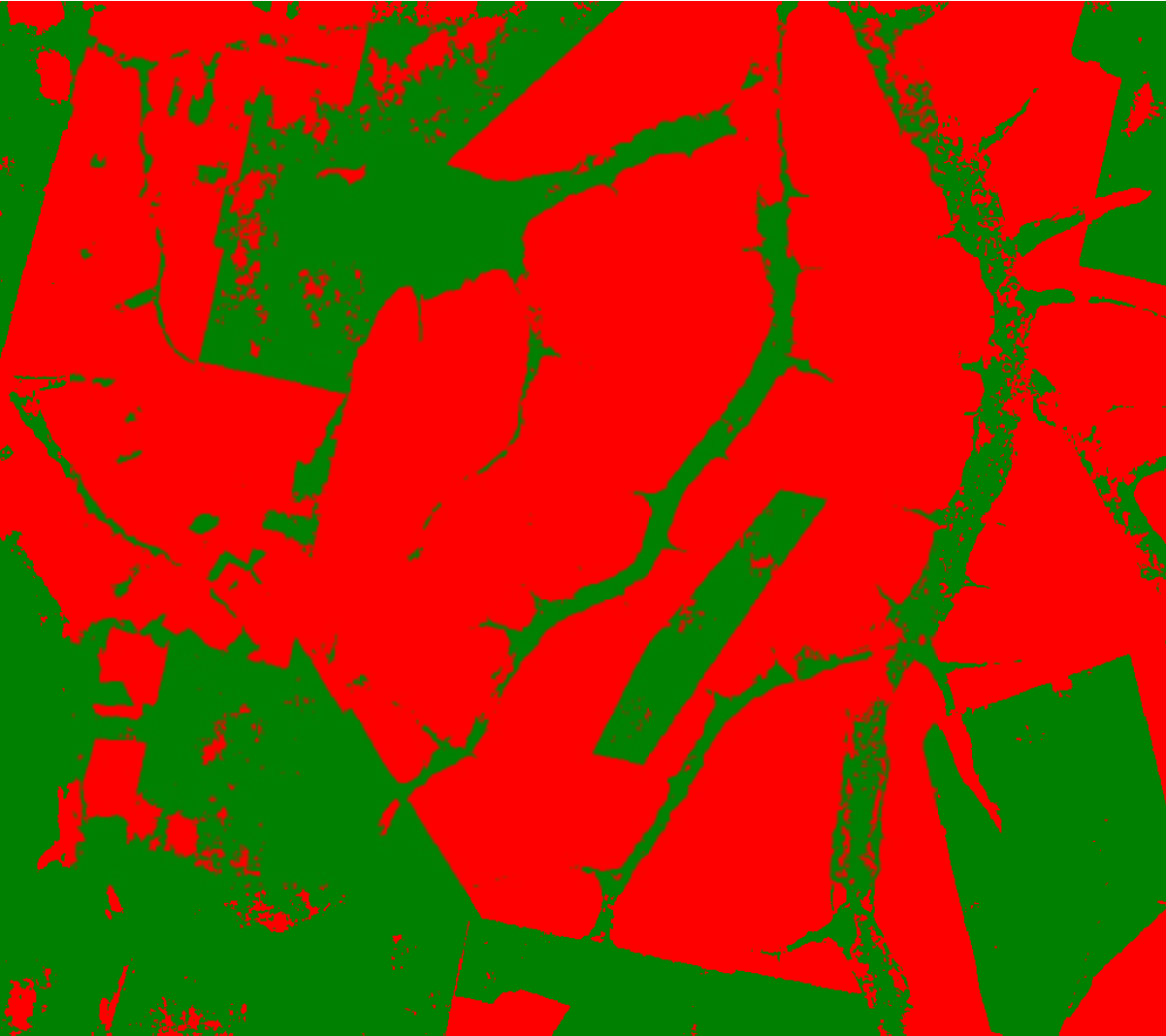}\\
		{(a)}&{(b)}&{(c)}&{(d)} \\
	\end{tabular}
	\caption{Prediction results from the proposed methods over the test areas. (a) Planet scope image acquired over the testing area. (b) Baseline Unet-like model trained on dense labels (c)  Incomplete supervision weakly supervised learning (d) Inaccurate supervision weakly supervised learning.   }\label{fig:Fig3}
\end{figure*}

\subsection{Dataset}

\noindent The proposed weakly supervised approaches were trained and tested using an analysis ready Sentinel-1 SAR ground range detected images in both the VV and VH polarizations \cite{mullissa2021sentinel} in Mato grosso state in Brazil that is characterized by a multi-layer evergreen forest structure at a nominal pixel size of 10m. The VV and VH channels were synthesized by taking the median value of 144 images acquired from January 1, 2019 to December 31, 2019.  The training area's upper left coordinates are $-53.83^0$ lon, $-9.70^0$ lat and the lower right coordinates are $-53.42^0$ lon, $-10.04^0$ lat. Whereas the testing area's upper left coordinates are $-52.59^0$ lon, $-11.93^0$ lat and the lower right coordinates are $-52.23^0$ lon, $-12.25^0$ lat. The Sentinel-1 SAR sensor acquires data in C band for both the VV and VH polarizations. The reference pixels for the incomplete WSL was extracted from the intersection between the JAXA ALOS-PALSAR2 based global four class forest/non-forest map \cite{shimada2014new}, Hansen global forest cover and forest cover/ forest loss datasets \cite{hansen2016humid} and the Land Use Change Alerts \cite{mullissa2024luca} and was visually checked using Planet scope high resolution images whereas for the inaccurate WSL, we synthesized the forest cover for 2020 by subtracting the globally available Hansen forest loss dataset from 2000-2019 from the 2000 forest cover dataset without quality control \cite{hansen2016humid} \cite{hansen2013high}. \\

\subsection{Network training}
\noindent We synthesized the training tensor by randomly selecting $11,995$ Sentinel-1 SAR image patches of size $64\times64\times2$ and a corresponding $11,995$ label patches with size $64\times64\times1$. The two channels in the input data patch correspond with the the VV and VH polarizations. To validate the model, we extracted 1990 patches from a geographically disjointed area to validate the model during training. \\

 \noindent The network was trained using Adam optimization method. We initialized the parameters randomly using the improved Xavier initialization. To diversify the training data, we applied data augmentation consisting of random vertical flips and $20^0$ rotations. In both WSL approaches we applied 32 filters in the first convolution block and progressively increased the filters by a factor of 2, 4, 8, 16, 32 times in each convolution blocks respectively (Fig. 1). We also applied a learning rate of 0.001 in both WSL architectures.  During training a mini-batches of 16 samples and a weight decay factor of $5\times 10^{-4}$ was used.   \\

\section{Results}
  To evaluate the performance of the proposed two WSL approaches,  we applied the trained models to a test image that was extracted from an area that was geographically disjointed from the training data region. We evaluated the accuracy of the trained models by deriving precision, recall and F-1 score by using the test dataset. We also compared the proposed approaches against a Unet-like architecture with identical parameters trained on accurately annotated dense labels. In this regard, the proposed WSL approaches achieved high accuracy that was comparable to the baseline Unet-like model that was trained on dense labels (Table~\ref{tbl:Tab5}). This indicated the proposed approaches are able to achieve high accuracy in mapping forest covers in a tropical setting using a small number annotated data. This was also confirmed when visually comparing the predicted forest/non-forest maps with that of a high resolution Planet scope image (Figure~\ref{fig:Fig3}). \\

\begin{table}[tbh]
	\begin{center}
		\footnotesize
		\setlength\tabcolsep{2pt}
		\begin{tabular}{lccccccc} \hline
		Method&Class&	Precision & Recall & F-1 score  \\ \hline			
		Incomplete supervision WSL &Non-forest&	0.90 &  0.973  & 0.935\\
                                    &Forest&	0.97 &  0.899  & 0.933\\
		Inaccurate supervision WSL& Non-forest & 0.902 &  0.983  & 0.94\\ 
                                    & Forest & 0.982 &  0.892  & 0.934\\ 
        Baseline dense segmentation Unet & Non-forest & 0.94 &  0.979  & 0.958\\ 
                                    & Forest & 0.978 &  0.891  & 0.934\\ \hline
		\end{tabular}\\
	\end{center}
	\caption{The precision, recall and F-1 score of the proposed incomplete, inaccurate weakly supervised learning approaches. }\label{tbl:Tab5}
\end{table}

 \noindent Overall, the proposed approaches achieved high accuracies in detecting forest cover from two weakly supervised approaches using SAR images.\\ 
 
\section{Conclusions}

\noindent In this abstract, we demonstrated the use of SAR images and two weakly supervised learning approaches for the mapping forests and non-forest regions in a tropical setting. We demonstrated a masked Unet-like architecture that trained a model from sparsely annotated reference labels and iterative refinement of a pseudo-label synthesized from globally available noisy reference dataset.\\

\noindent In future work, we will further improve the efficacy of the method by minimizing overfitting by using a diverse set of training dataset acquired from different eco-regions. Furthermore, we will expand the prediction capability of the model from binary segmentation to a multi-class landcover segmentation problem. Finally, we'll explore different fusion approaches between multi-modal datasets to further improve the robustness of the proposed method.\\

\bibliographystyle{IEEEbib}
\bibliography{Reference.bib}

\begin{thebibliography}{10}

\bibitem{lewis2009changing}
Simon~L Lewis, Jon Lloyd, Stephen Sitch, Edward~TA Mitchard, and William~F
  Laurance,
\newblock ``Changing ecology of tropical forests: evidence and drivers,''
\newblock {\em Annual Review of Ecology, Evolution, and Systematics}, vol. 40,
  pp. 529--549, 2009.

\bibitem{hansen2013high}
Matthew~C Hansen, Peter~V Potapov, Rebecca Moore, Matt Hancher, Svetlana~A
  Turubanova, Alexandra Tyukavina, David Thau, Stephen~V Stehman, Scott~J
  Goetz, Thomas~R Loveland, et~al.,
\newblock ``High-resolution global maps of 21st-century forest cover change,''
\newblock {\em science}, vol. 342, no. 6160, pp. 850--853, 2013.

\bibitem{kussul2017deep}
Nataliia Kussul, Mykola Lavreniuk, Sergii Skakun, and Andrii Shelestov,
\newblock ``Deep learning classification of land cover and crop types using
  remote sensing data,''
\newblock {\em IEEE Geoscience and Remote Sensing Letters}, vol. 14, no. 5, pp.
  778--782, 2017.

\bibitem{ronneberger2015u}
Olaf Ronneberger, Philipp Fischer, and Thomas Brox,
\newblock ``U-net: Convolutional networks for biomedical image segmentation,''
\newblock in {\em Medical Image Computing and Computer-Assisted
  Intervention--MICCAI 2015: 18th International Conference, Munich, Germany,
  October 5-9, 2015, Proceedings, Part III 18}. Springer, 2015, pp. 234--241.

\bibitem{zhou2018brief}
Zhi-Hua Zhou,
\newblock ``A brief introduction to weakly supervised learning,''
\newblock {\em National science review}, vol. 5, no. 1, pp. 44--53, 2018.

\bibitem{mullissa2023deep}
Adugna Mullissa, Johannes Reiche, and Martin Herold,
\newblock ``Deep learning and automatic reference label harvesting for
  sentinel-1 sar-based rapid tropical dry forest disturbance mapping,''
\newblock {\em Remote Sensing of Environment}, vol. 298, pp. 113799, 2023.

\bibitem{wang2020weakly}
Sherrie Wang, William Chen, Sang~Michael Xie, George Azzari, and David~B
  Lobell,
\newblock ``Weakly supervised deep learning for segmentation of remote sensing
  imagery,''
\newblock {\em Remote Sensing}, vol. 12, no. 2, pp. 207, 2020.

\bibitem{li2023one}
Zhenshi Li, Xueliang Zhang, and Pengfeng Xiao,
\newblock ``One model is enough: Toward multiclass weakly supervised remote
  sensing image semantic segmentation,''
\newblock {\em IEEE Transactions on Geoscience and Remote Sensing}, 2023.

\bibitem{li2021effectiveness}
Zhenshi Li, Xueliang Zhang, Pengfeng Xiao, and Zixian Zheng,
\newblock ``On the effectiveness of weakly supervised semantic segmentation for
  building extraction from high-resolution remote sensing imagery,''
\newblock {\em IEEE Journal of Selected Topics in Applied Earth Observations
  and Remote Sensing}, vol. 14, pp. 3266--3281, 2021.

\bibitem{shimada2014new}
Masanobu Shimada, Takuya Itoh, Takeshi Motooka, Manabu Watanabe, Tomohiro
  Shiraishi, Rajesh Thapa, and Richard Lucas,
\newblock ``New global forest/non-forest maps from alos palsar data
  (2007--2010),''
\newblock {\em Remote Sensing of environment}, vol. 155, pp. 13--31, 2014.

\bibitem{mullissa2024luca}
Adugna Mullissa, Sassan Saatchi, Ricardo Dalagnol, Tyler Erickson, Naomi
  Provost, Fiona Osborn, Aleena Ashary, Violet Moon, and Daniel Melling,
\newblock ``Luca: A sentinel-1 sar-based global forest land use change alert,''
\newblock {\em Remote Sensing}, vol. 16, no. 12, pp. 2151, 2024.

\bibitem{mullissa2021sentinel}
Adugna Mullissa, Andreas Vollrath, Christelle Odongo-Braun, Bart Slagter,
  Johannes Balling, Yaqing Gou, Noel Gorelick, and Johannes Reiche,
\newblock ``Sentinel-1 sar backscatter analysis ready data preparation in
  google earth engine,''
\newblock {\em Remote Sensing}, vol. 13, no. 10, pp. 1954, 2021.

\bibitem{hansen2016humid}
Matthew~C Hansen, Alexander Krylov, Alexandra Tyukavina, Peter~V Potapov,
  Svetlana Turubanova, Bryan Zutta, Suspense Ifo, Belinda Margono, Fred Stolle,
  and Rebecca Moore,
\newblock ``Humid tropical forest disturbance alerts using landsat data,''
\newblock {\em Environmental Research Letters}, vol. 11, no. 3, pp. 034008,
  2016.

\end{thebibliography}

\end{document}